\documentclass[10pt,letterpaper,twocolumn,english,aps,showpacs]{revtex4} 

\usepackage[T1]{fontenc}
\usepackage[latin9]{inputenc}
\usepackage{textcomp}
\usepackage{amstext}
\usepackage{graphicx}
\usepackage{esint}

\makeatletter

\pdfpageheight\paperheight
\pdfpagewidth\paperwidth

\newcommand{\lyxmathsym}[1]{\ifmmode\begingroup\def\b@ld{bold}
  \text{\ifx\math@version\b@ld\bfseries\fi#1}\endgroup\else#1\fi}

\@ifundefined{textcolor}{}
{%
 \definecolor{BLACK}{gray}{0}
 \definecolor{WHITE}{gray}{1}
 \definecolor{RED}{rgb}{1,0,0}
 \definecolor{GREEN}{rgb}{0,1,0}
 \definecolor{BLUE}{rgb}{0,0,1}
 \definecolor{CYAN}{cmyk}{1,0,0,0}
 \definecolor{MAGENTA}{cmyk}{0,1,0,0}
 \definecolor{YELLOW}{cmyk}{0,0,1,0}
 }

\makeatletter
\gdef\@ptsize{0}
\let\@currsize\normalsize
\makeatother
\usepackage{setspace}
\makeatother

\usepackage{babel}

\newcommand{\bea}{\begin{eqnarray}}
\newcommand{\eea}{\end{eqnarray}}
\newcommand{\be}{\begin{equation}}
\newcommand{\ee}{\end{equation}}
\newcommand{\LL}{\mathcal{L}}
\newcommand{\tr}{{\rm tr}}
\newcommand{\nn}{\nonumber}

\begin{document}

\title{The nature of self-localization of Bose-Einstein condensates in deep optical lattices}

\author{Holger Hennig$^{1}$, Ragnar Fleischmann$^{2}$}

\affiliation{$^{1}$Department of Physics, Harvard University, Cambridge, MA 02138,
USA}

\affiliation{$^{2}$Max Planck Institute for Dynamics and Self-Organization, 37073
G\"ottingen, Germany}
\begin{abstract}
We analyze the nature of a novel type of self-trapping transition called
self-localization (SL) of Bose-Einstein condensates in one-dimensional
optical lattices in the presence of weak local dissipation. SL
has recently been observed in several studies based upon the discrete
nonlinear Schr\"odinger equation (DNLS),
however, its origin is hitherto an open question. We show that SL
is based upon a self-trapping crossover in the system.
Furthermore, we establish
that the origin of the crossover is the Peierls-Nabarro barrier,
an energy threshold describing the stability of self-trapped states.
Beyond the mean-field description the crossover becomes even sharper which is also reflected
by a sudden change of the coherence of the condensate. While we expect that the crossover
can be readily studied in current experiments in deep optical lattices,
our results allow for the preparation of robust and long-time coherent
quantum states.
\end{abstract}

\maketitle

\section{Introduction}
Dissipation is typically known to represent a major obstacle in the
coherent control of quantum systems. However, in recent years, a strong
interest in engineered dissipation has evolved, where dissipation
has been used as a tool for quantum state preparation \cite{Diehl:2008ha,Kraus:2008jd}
as well as quantum information processing and entanglement generation
\cite{Verstraete:2009kc} and to induce self-trapping (ST) \cite{Gericke:2008tq,Graefe:2008cl,Witthaut:2011kv,Trimborn:2011gy}.
Bose-Einstein condensates (BECs) have been shown to support a variety of different kinds of ST,
both in the continuous case (such as bright and dark solitons \cite{Burger:1999vs,Khaykovich:2002tc,Strecker:2002vg,Eiermann:2004vt,Cornish:2006ii,Stellmer:2008km})
and in discrete systems \cite{Albiez:2005wk,Zibold:2010el,Rasmussen:2000ur,Rasmussen:2000vn,Raghavan:1999ud,Trombettoni:2001wl,Trombettoni:2001tb,Smerzi:1997vd,Hennig:2010gy}.
A particularly high level of control has been achieved in a two-mode
BEC \cite{Zibold:2010el}, where ST can also be induced by local dissipation which can even repurify a BEC \cite{Witthaut:2008un}.

A novel self-trapping transition coined `self-localization' has been
observed numerically in several studies based upon the DNLS in the
presence of weak boundary dissipation in one-dimensional 
deep optical lattices
\cite{Livi:2006id,Franzosi:2007ds,Ng:2009tu,Franzosi:2011ft}. In contrast to self-trapping,
where a system is either prepared in a self-trapped state \cite{Albiez:2005wk,Zibold:2010el,Franzosi:2010fz}
or driven towards it \cite{Witthaut:2011kv,Trimborn:2011gy,Gericke:2008tq}, SL is a mechanism where in presence of weak local or boundary dissipation a very general initially diffusive state leads to the formation of one or more discrete breathers (DBs, see \cite{Flach:2008ud,Campbell:2004vz}
for an overview). 
However, SL was only found, if the atomic interaction strength exceeds a critical value  \cite{Ng:2009tu}. While the phenomenology of SL has been studied, the mechanisms that lead to this transition have remained unknown up to now.

In this letter, we propose a mechanism for SL allowing us to give an explicit formula for an upper bound estimate of the SL threshold for the DNLS in excellent agreement with the numerical findings of \cite{Ng:2009tu}. 
The mechanism is based on a `crossover' which surprisingly becomes much sharper when quantum corrections beyond the mean-field description are included, which is observed, e.g., in the condensate fraction of the system.
Our work also contributes to clarify conditions for the experimental observation of SL, as discussed at the end of the article.

To understand the nature of SL it is essential to note that the fixed point corresponding to the DB state into which the initial condition collapses does not undergo a bifurcation itself. On the contrary, using standard methods  \cite{Carr:1985ui,Aubry:1997wo,Darmanyan:1998tw,Proville:1999vb} the bright breather fixed point can easily be numerically found to exist and to be linearly stable for \emph{all} positive nonlinearity strengths.
Linear stability analysis therefore does not suffice to understand the SL transition.
The underlying idea of our approach is that near the SL threshold a single strong, localized fluctuation of the number of atoms locally brings the system's state into the basin of attraction of a DB fixed point in phase space. The role of dissipation at this point is that DBs are attractors in dissipative systems \cite{Flach:1998ul,Martnez:2003bw,MacKay:1998um}, while Hamiltonian systems do not have attractors. 
In the simplest and most likely event a strong increase in the number of atoms happens on a single site that will become the center of the DB to be formed. 

We therefore study first, how a single site excitation can lead to the formation of a DB and find that there exists a distinct nonlinearity strength at which this initial condition crosses over into a self-trapped state. 
We show that the origin of this \emph{ST crossover} is an energy threshold describing the stability of self-trapped states (called the Peierls-Nabarro (PN) energy barrier \cite{Kivshar:1993vf,Rumpf:2004en,Hennig:2010gy}).
Secondly, we statistically estimate the critical nonlinearity at the onset of SL by studying the probability that a fluctuation in a diffusive state exceeds this ST crossover and leads to the formation of a breather.
The ST crossover and SL should not only be observable for BECs but as well, e.g., in coupled nonlinear optical waveguides \cite{Christodoulides:2003vu,Flach:2008ud}.

\begin{figure}
\centering
\includegraphics[width=0.9\columnwidth]{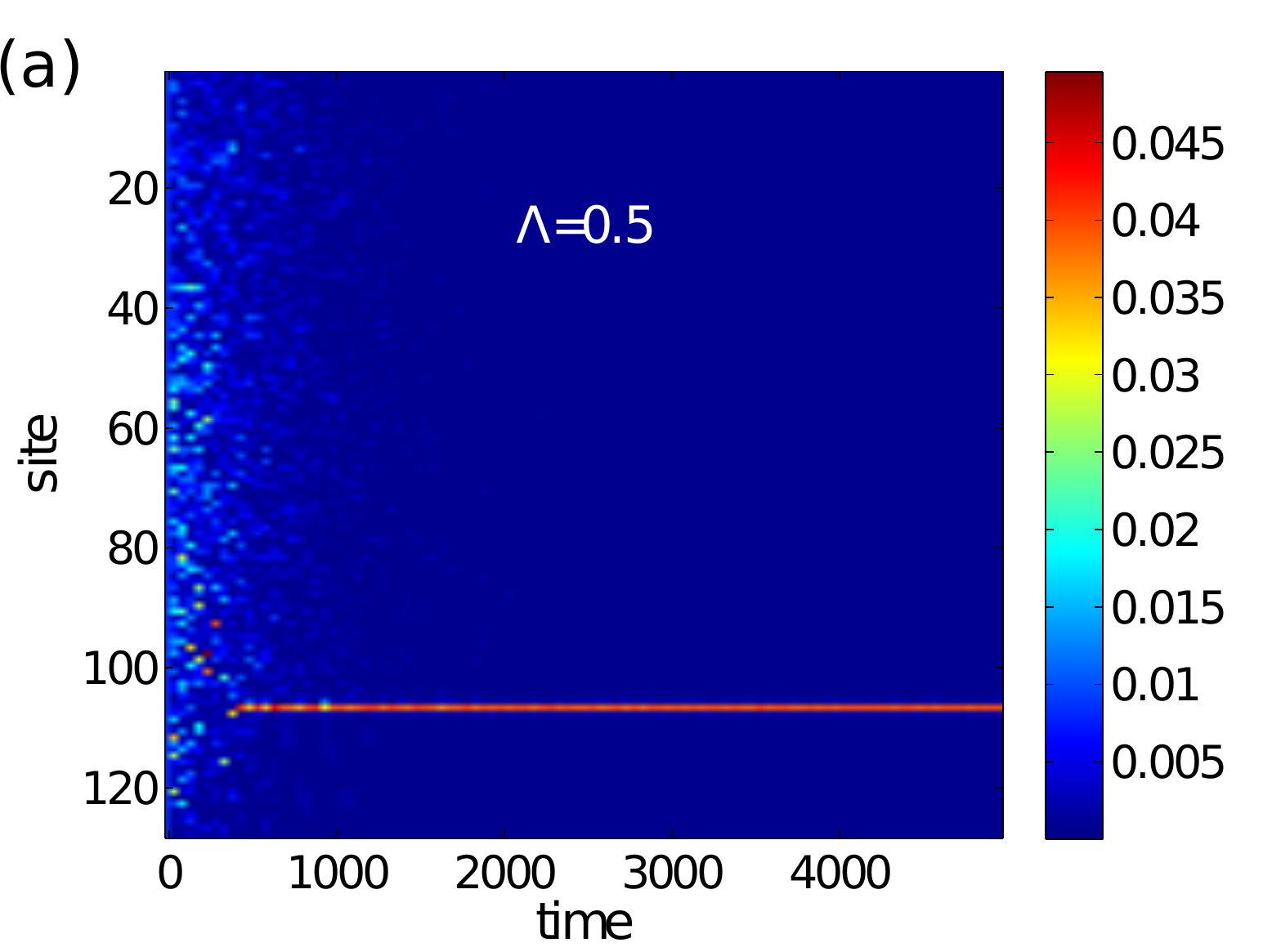}
\includegraphics[width=0.9\columnwidth]{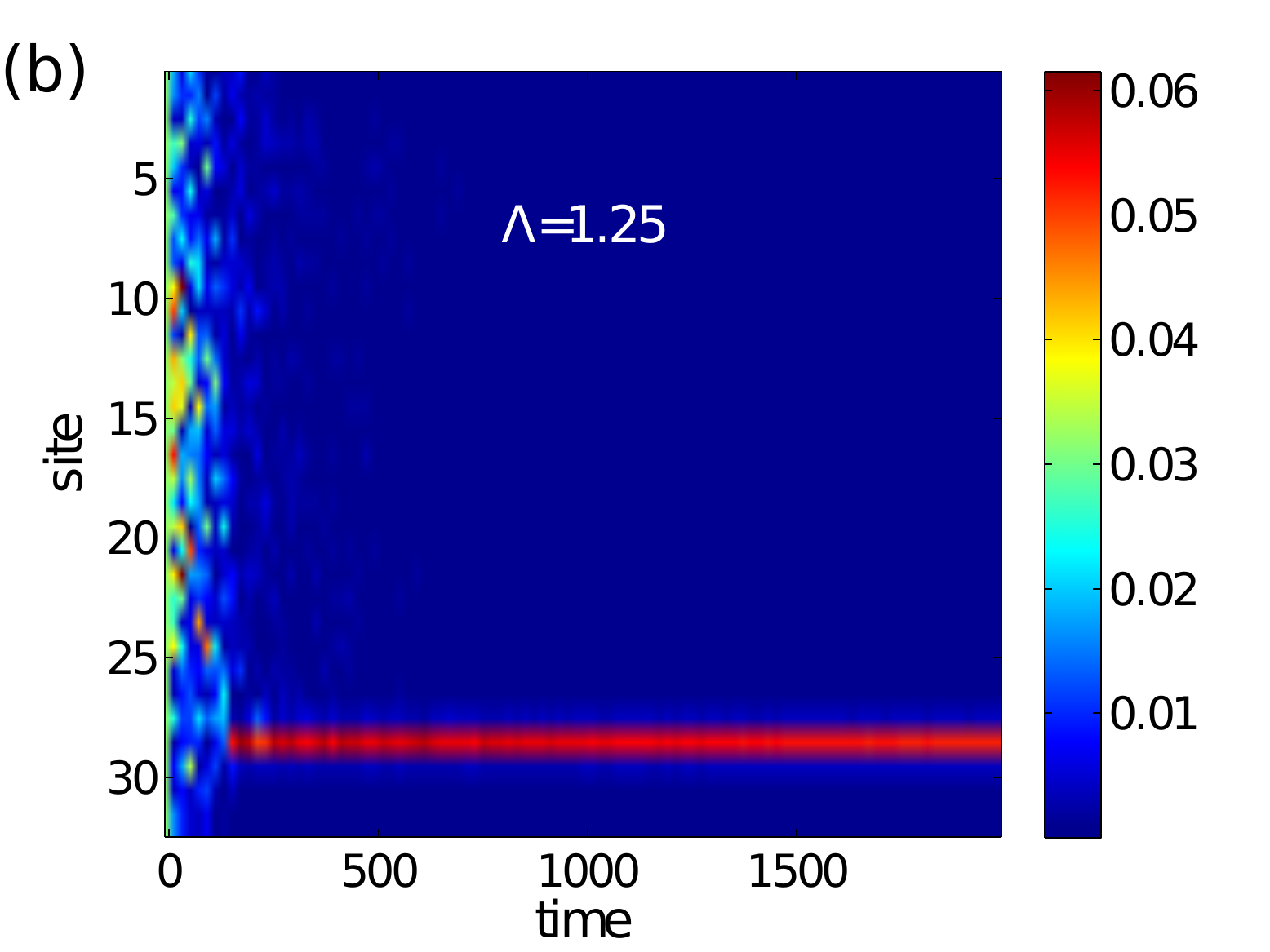}
\caption{Demonstration of self-localization in a lattice with {\bf (a)} $M=128$ and {\bf (b)} $M=32$ sites based upon the dissipative DNLS. 
The color code shows $|\psi_{n}(t)|^{2}$ (normalized to $1$ at $t=0$). Time is measured in units of the tunneling rate J. The initial condition is a homogeneously populated lattice (i.e., constant norms) with random phases at each site uniformly drawn from $[0,2\pi]$. Boundary dissipation rate at sites $1$ and $M$ is $\gamma=0.2$.
{\bf (a)} The dissipative dynamics leads to formation of a discrete breather centered at site $107$.
The effective nonlinearity $\Lambda=L/M$ is $\Lambda=0.5$ just above the self-localization threshold $\Lambda_b=0.58$ (see Eq.~\ref{eq:Lambdac_final}).
{\bf (b)} A discrete breather forms centered at site $29$. The effective nonlinearity is $\Lambda=1.25$. In both panels, $\Lambda$ is larger than the self-localization threshold $\Lambda_b$ (cf.~Eq.~\ref{eq:Lambdac_final}). Note that though the discrete breathers are more likely to form near the middle of the lattice, they can also emerge near the boundaries, as shown here.
\label{fig:sl}}
\end{figure}

Consider the Bose-Hubbard Hamiltonian in the mean-field
description \cite{PETHICK:2008tn,Buonsante:2008fe}
\begin{equation}
H=U\sum_{i=n}^{M}|\psi_{n}|^{4}-\frac{J}{2}\sum_{n=1}^{M-1}(\psi_{n}^{*}\psi_{n+1}+\text{c.c.)}\label{eq:hamiltonian}
\end{equation}
with on-site interaction $U$, tunneling rate $J$, lattice index
$n=1\ldots M$, where $M$ denotes the number of lattice sites. Including
boundary dissipation, the mean-field equations of motion are given by the dissipative
DNLS (see \cite{Jeffers:2000ti,Witthaut:2011kv,Trimborn:2011gy} for a derivation
of the loss term) 
\begin{equation}
i\dot{\psi}_{n}=L|\psi_{n}|^{2}\psi_{n}-\frac{1}{2}(\psi_{n-1}+\psi_{n+1}\!)-i\gamma \psi_{n}(\delta_{n,1}+\delta_{n,M}\!) \label{eq:dnls}
\end{equation}
with $\hbar=1$, nonlinearity $L=(2U/J){\cal{N}}$, 
dissipation rate $\gamma$, total number of atoms ${\cal{N}}$ and the normalization $\sum_{n}|\psi_{n}|^{2}=1.$ We introduce a measure of the \textit{local nonlinearity} 
$
L_{n}^{\text{local}}=(2U{\cal{N}}/J) N_{n}\label{eq:Llocal}
$
, where $N_{n}=|\psi_{n}|^{2}$ is the relative number of atoms (also referred
to as the norm) at site n.

\section{self-localization vs.~self-trapping}
Though SL is based upon ST, it is distinguished by the way in which a stable (or metastable) and spatially localized state is reached. There are several ways to obtain self-trapping of BECs in optical
lattices which we classify into three types.

Type I (\textit{`static preparation'}): The quantum system is prepared in (or sufficiently
close to) a self-trapped state. This has been realized in various experiments \cite{Eiermann:2004vt,Albiez:2005wk,Bloch:2005uv,Zibold:2010el}.
Using a variational approach, a phase diagram has been calculated, that describes the transition
from diffusion to ST for an initial Gaussian wave packet \cite{Trombettoni:2001wl,Trombettoni:2001tb}, which, however, does not account for SL. Note that recent numerics for the DNLS \cite{Franzosi:2011ft} rather contradicts the phase diagram in \cite{Trombettoni:2001wl}.

Type II (\textit{`dynamical preparation'}): Another route to ST is to apply
a strong local dissipation pulse, which can depopulate one or more sites and create a stable
isolated peak or vacancy \cite{Gericke:2008tq,Witthaut:2008un,Witthaut:2011kv,Trimborn:2011gy}
(leading to the formation of a bright or dark breather). In particular, spatially
resolved dissipative manipulation in an optical lattice
using an electron beam with single-site addressability has been demonstrated
\cite{Gericke:2008tq}.

\begin{figure}
\centering
\includegraphics[width=1\columnwidth]{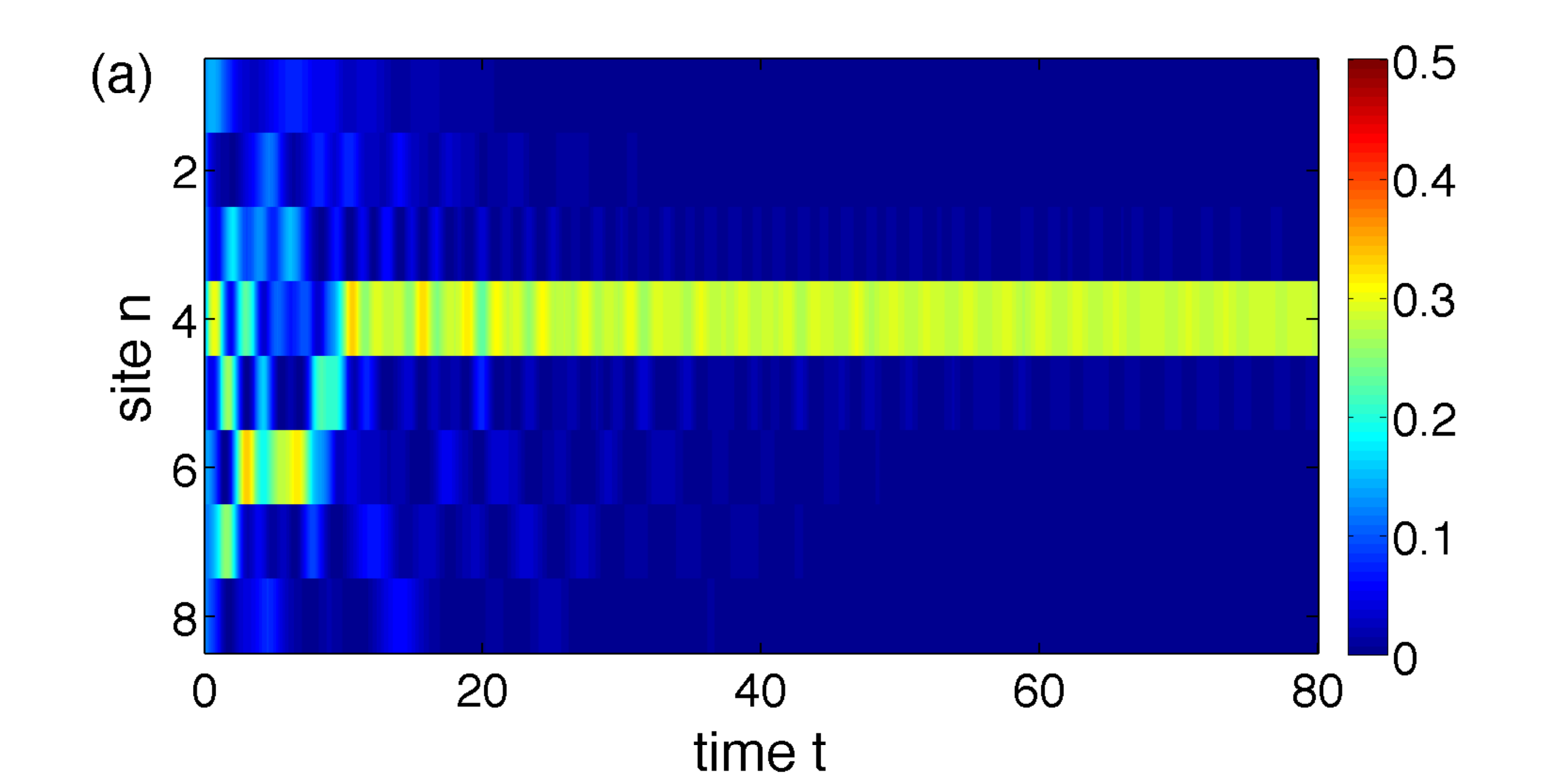}
\includegraphics[width=1\columnwidth]{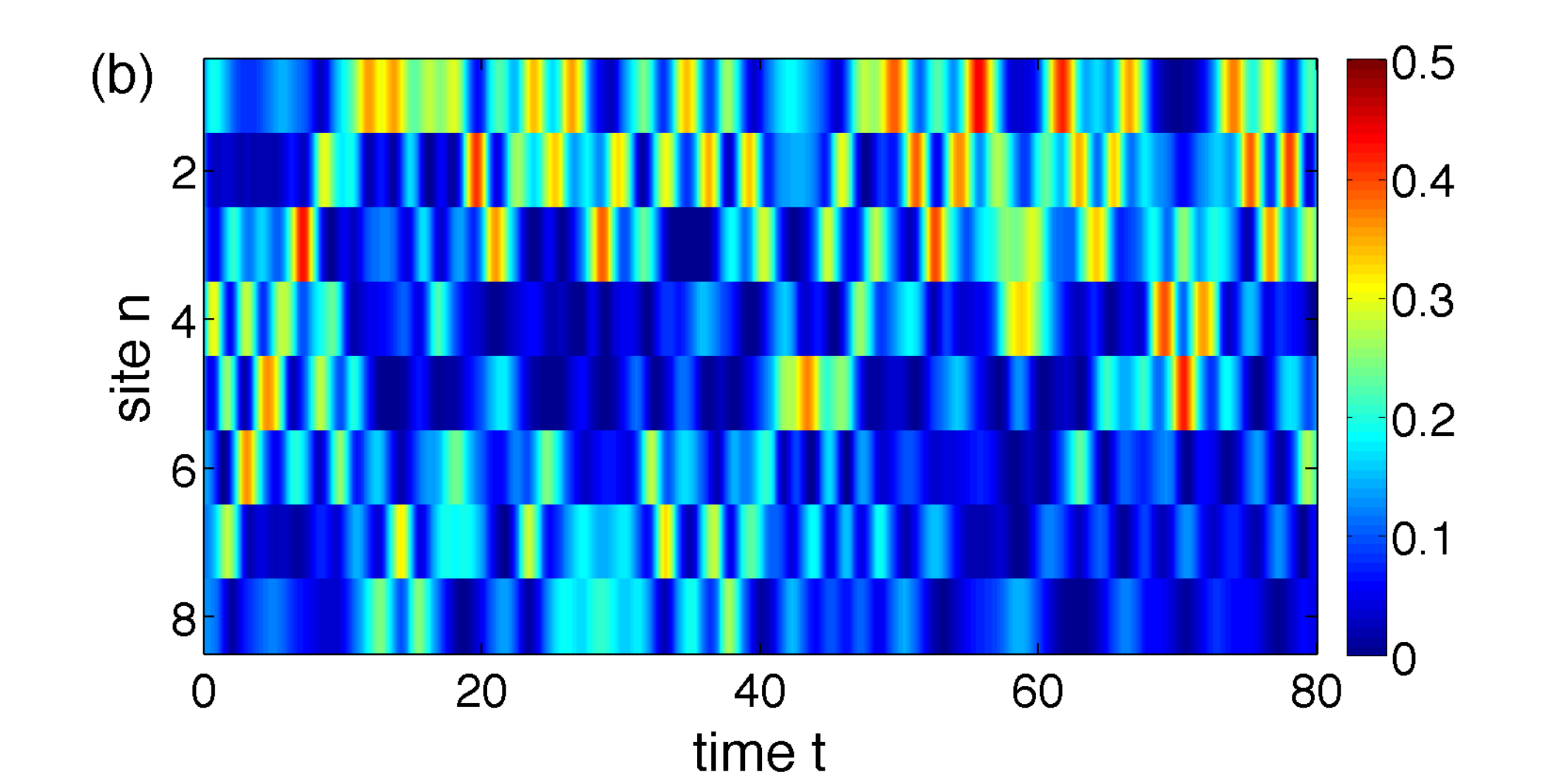}
\includegraphics[width=1\columnwidth]{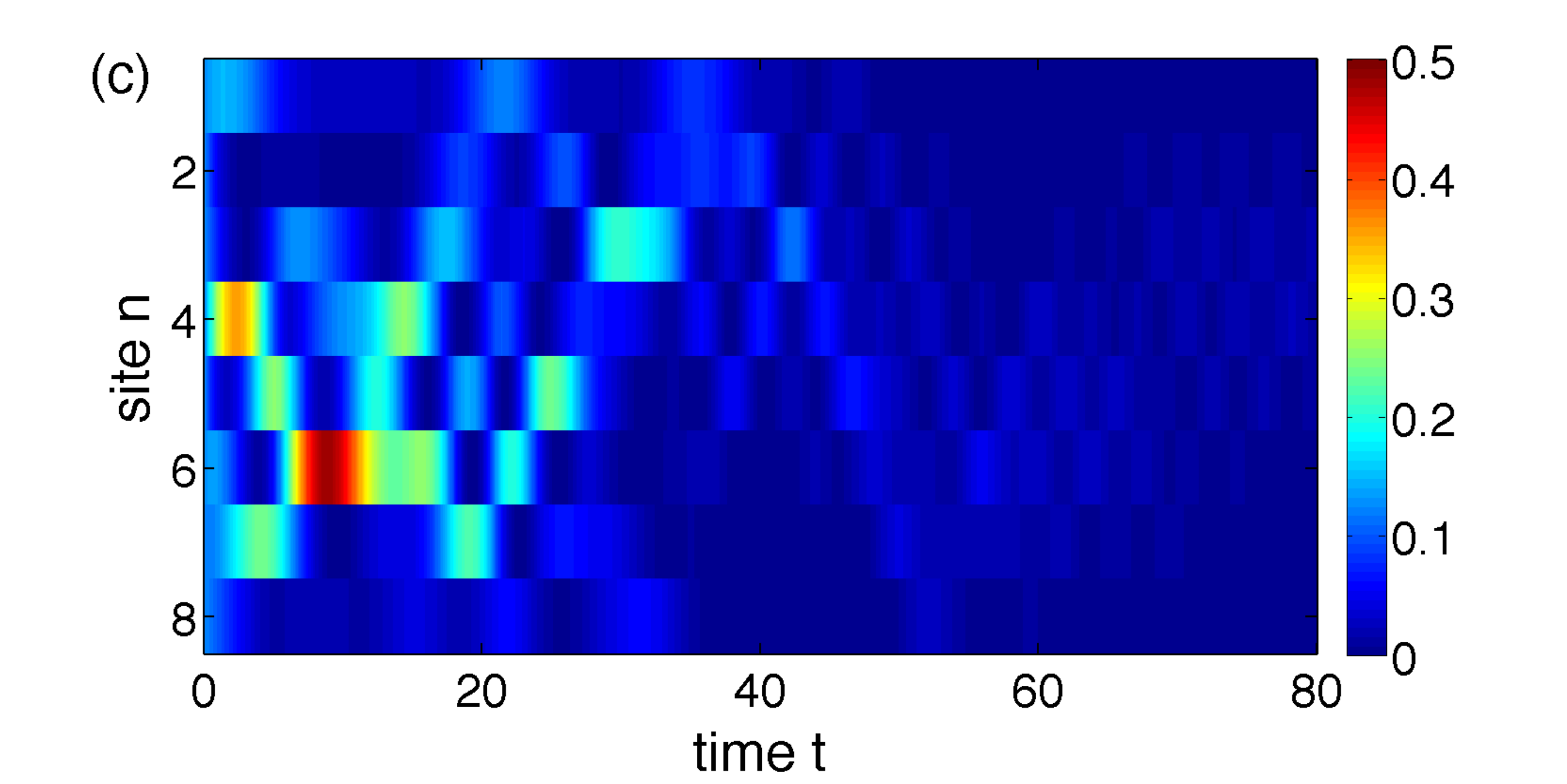}
\caption{Self-localization can also be observed in small lattices, shown here for $M=8$ wells (similar to the experimental setup in \cite{Esteve:2008vr}) for the DNLS with boundary dissipation rate $\gamma=0.2$. The color code shows the atomic density $|\psi_{n}(t)|^{2}$. Initial condition is a homogeneously populated lattice with random phases as in Fig.~\ref{fig:sl}. Three representative cases with identical initial condition are shown, SL is observed only in panel (a). Time is measured in units of $J$, i.e.~for $J=10$Hz, which is a typical experimental value in \cite{Esteve:2008vr}, the discrete breather in (a) forms around time $t=1$s.
{\bf (a)} Self-localization with $\Lambda=1.354=1.5\Lambda_b$.
{\bf (b)} Without dissipation ($\gamma=0$) no self-localization takes place. For times $t\gtrsim 5/J$, the plot strongly differs qualitatively from the dissipative case (a). 
We carefully checked for different parameter regimes $M$ and $\Lambda$, that even for times several orders of magnitude longer than depicted here, SL does not occur for $\gamma=0$. 
{\bf (c)} For the dissipative case where $\Lambda=0.5<\Lambda_b$, no self-localization takes place and the number of atoms in the lattice decays quickly.
\label{fig:sl2}}
\end{figure}

Type III (\textit{`self-localization'}): A third way to generate self-trapping is SL, where the
system prepared in a random (generic) state in the presence of  boundary or other local dissipation dynamically forms one or more DBs, see Figs.~\ref{fig:sl} and \ref{fig:sl2}. In contrast to Type II, the positions where DBs form are not determined by the location of the leak  \cite{Livi:2006id,Franzosi:2007ds,Ng:2009tu,Franzosi:2011ft}. In absence of boundary or local dissipation SL does not take place, see Fig.~\ref{fig:sl2}(b) and \cite{Livi:2006id}. Also, below a threshold $\Lambda_b$, self-localization does not occur (cf.~Fig.~\ref{fig:sl2}(c)). This threshold to SL has been observed in detail in ref.~\cite{Ng:2008tu}, in particular for lattices with large number of wells ($M=128$ to $M=4096$). A main purpose of this article is to derive an explicit formula for the SL threshold $\Lambda_b$ (cf.~Eq.~\ref{eq:Lambdac_final}).
%

\section{Self-trapping crossover}
\begin{figure}
\centering\includegraphics[width=1\columnwidth]{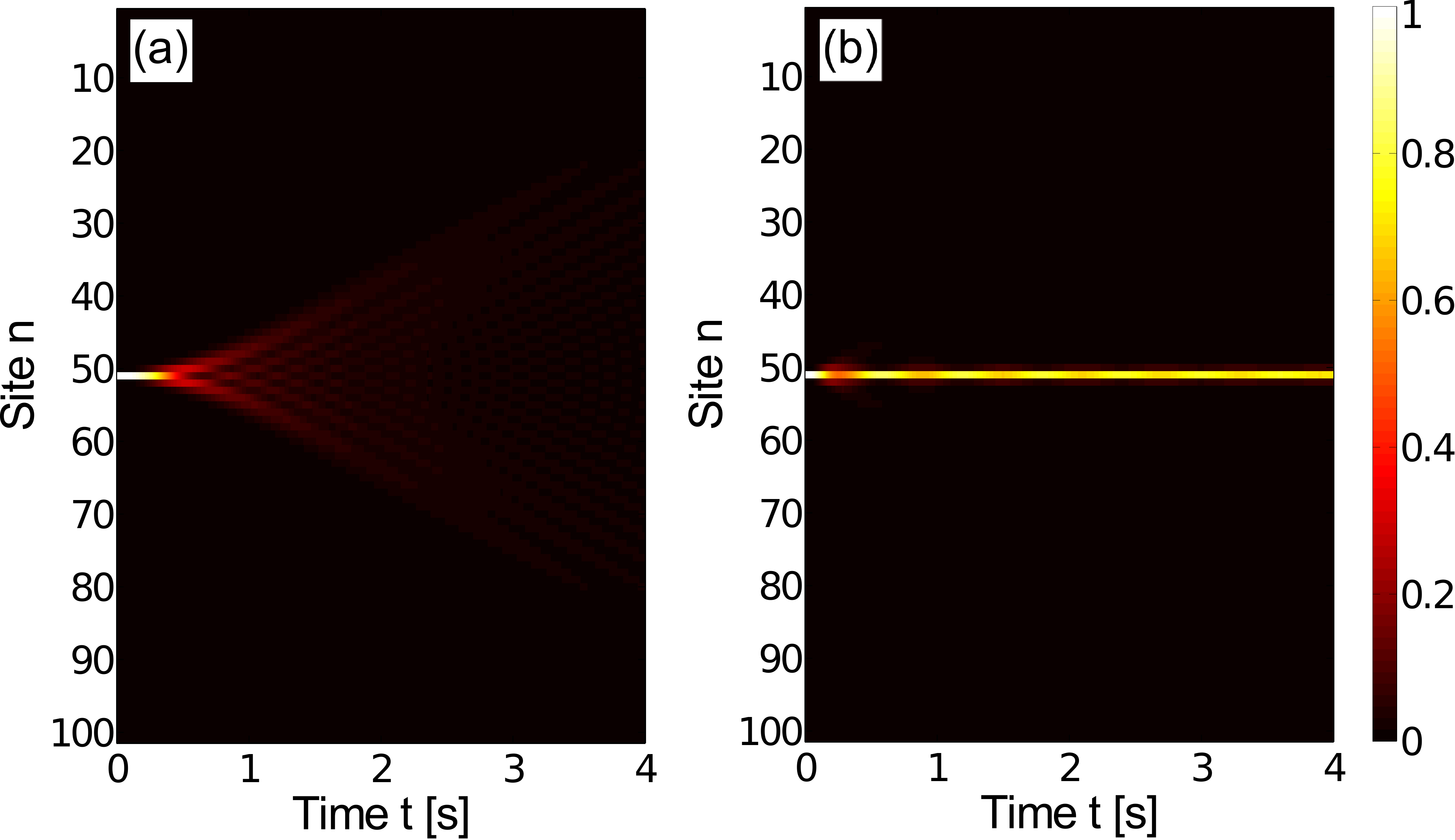}
\caption{A ST crossover for a $\delta$-like initial condition,
where all atoms are located at a single site $c$, is found both beyond
and within the mean-field description (depicted here for the DNLS).
The color code shows the normalized atomic density $|\psi_{n}(t)|^{2}$.
\textbf{(a)} Below the crossover, for $L=1.6<L_{\text{\text{co}}}$,
the localized peak at $t=0$ decays exponentially fast. \textbf{(b)}
Above the crossover (shown is $L=2.4>L_{\text{co}}$) a discrete
breather forms. The particle density is stable and decays exponentially
in space away from the center. About $85$\% of the atoms are located
in three sites after time $t=4\,$s. Other parameters are $M=101$,
$c=51$, $J=10\,$Hz and $\gamma=0$. \label{fig:crossover}}
\end{figure}

\begin{figure}[!ht]
\includegraphics[width=0.48\columnwidth]{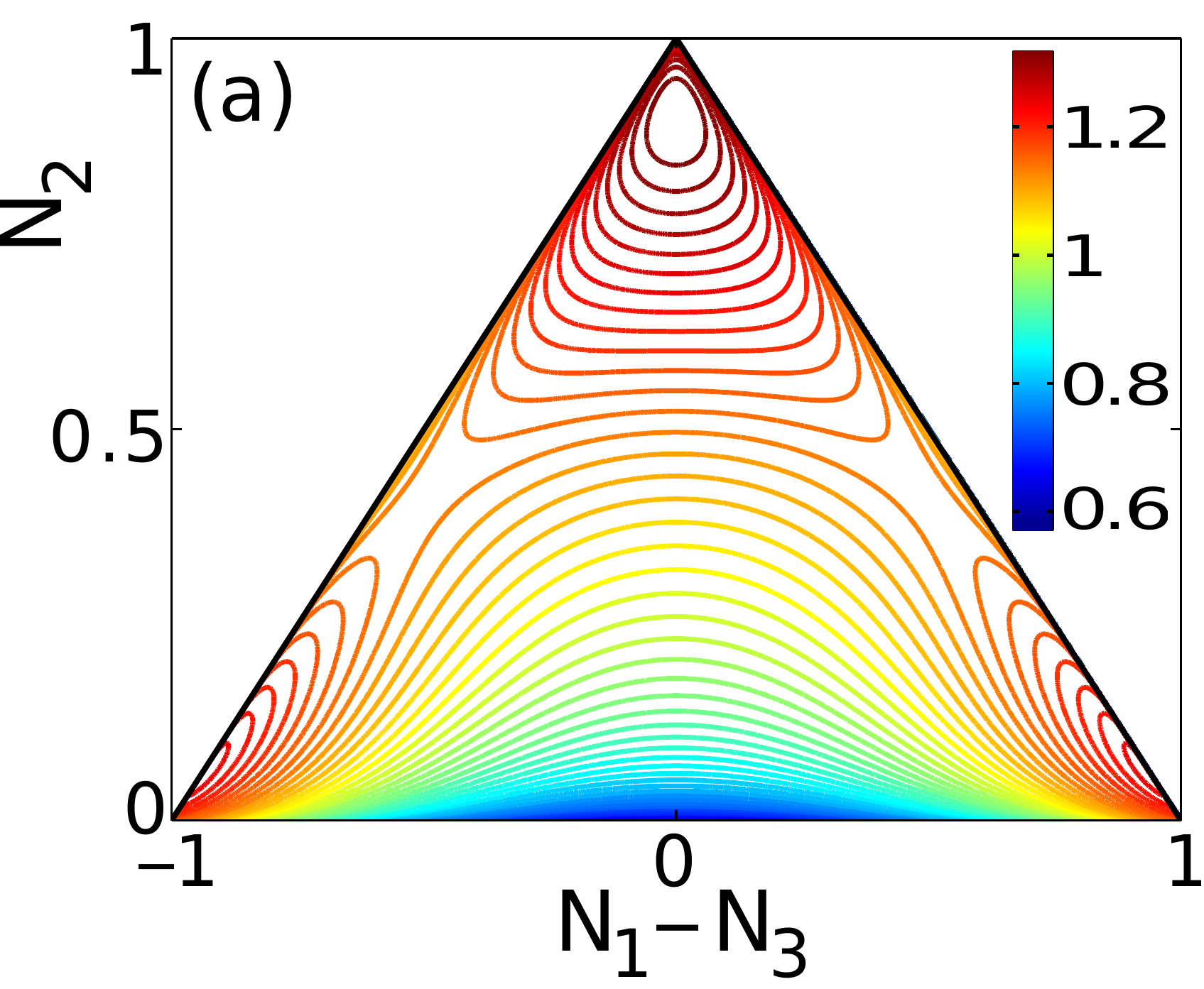}\enskip{}\includegraphics[width=0.48\columnwidth]{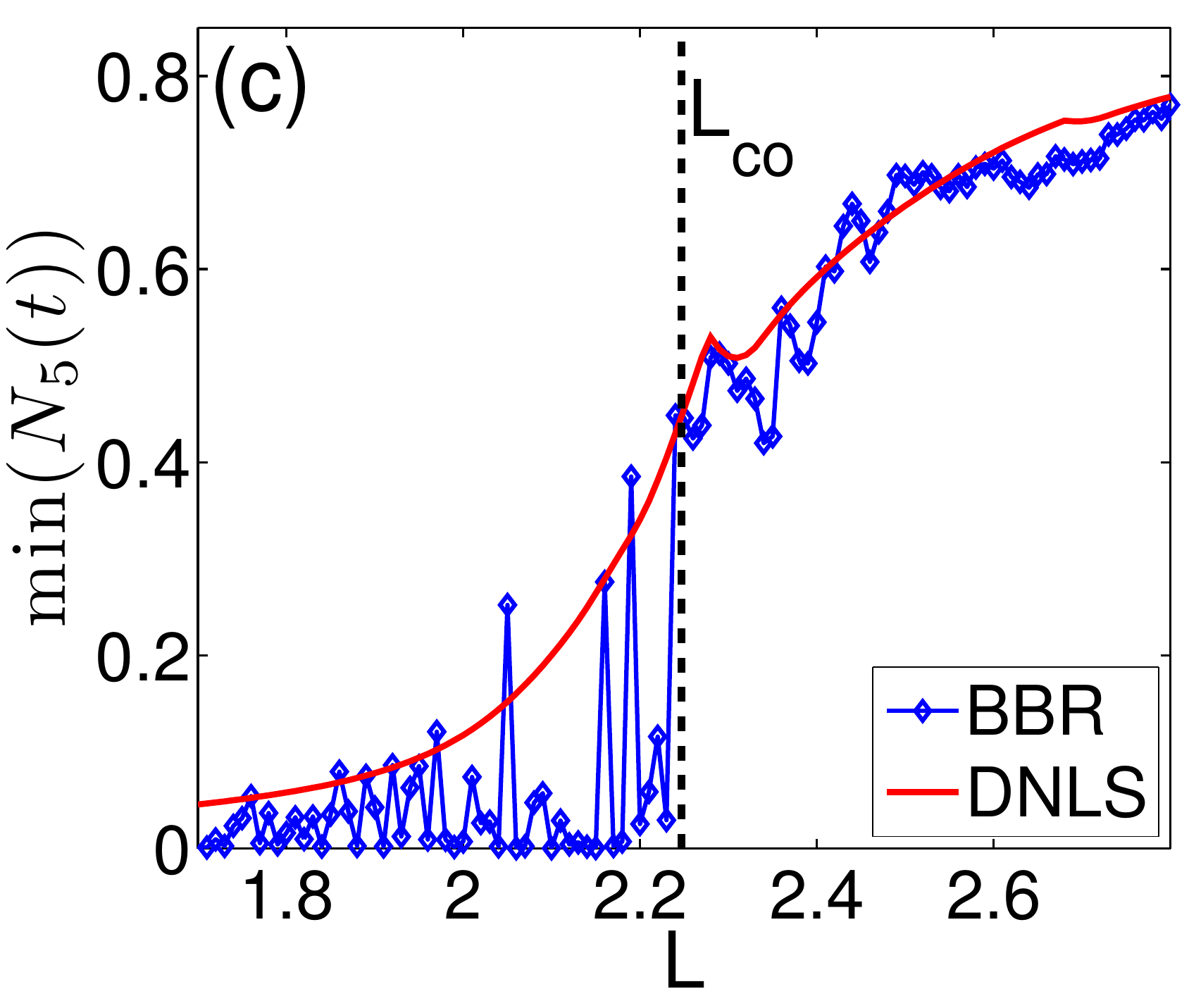}
\includegraphics[width=0.48\columnwidth]{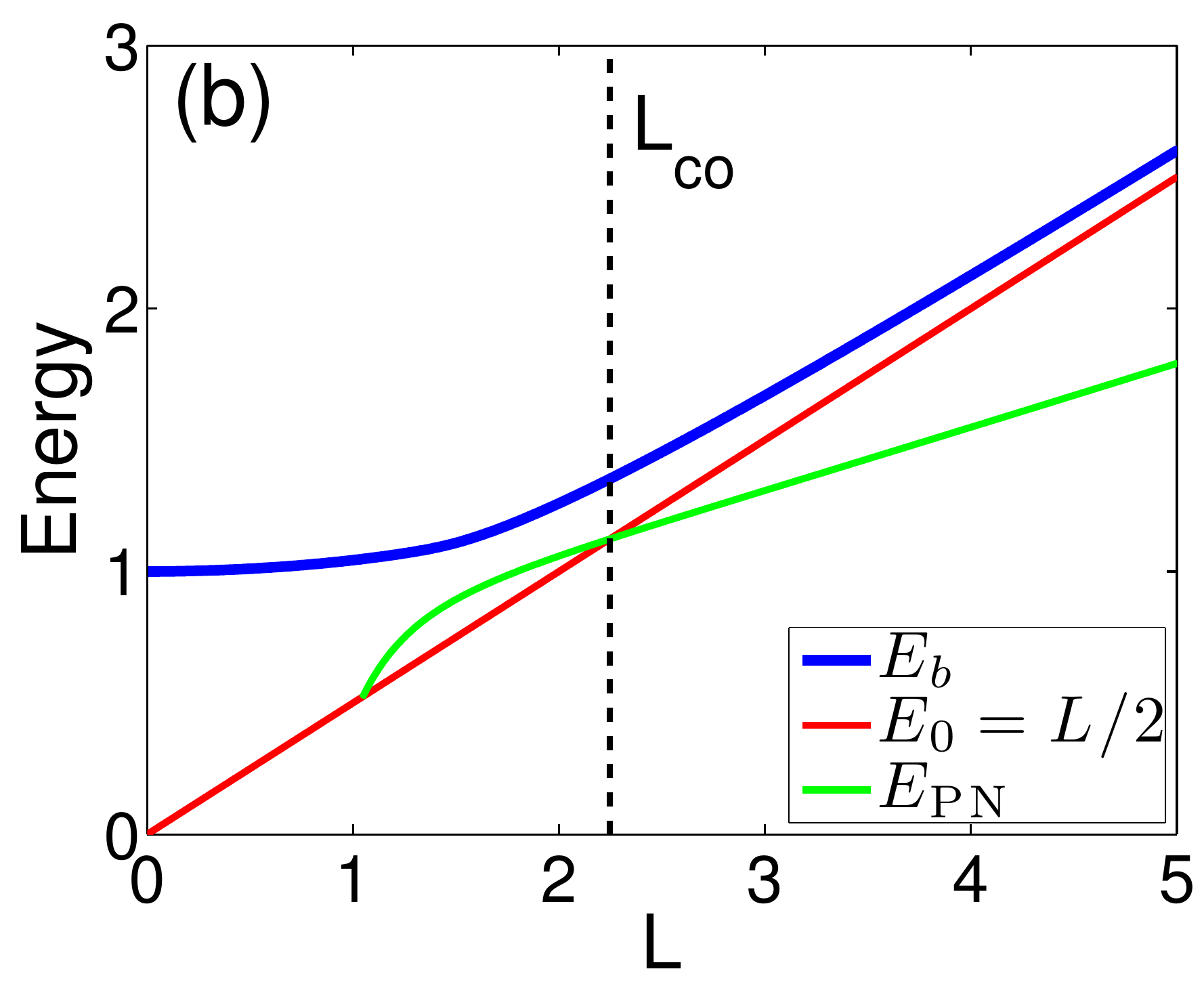}\enskip{}\includegraphics[width=0.48\columnwidth]{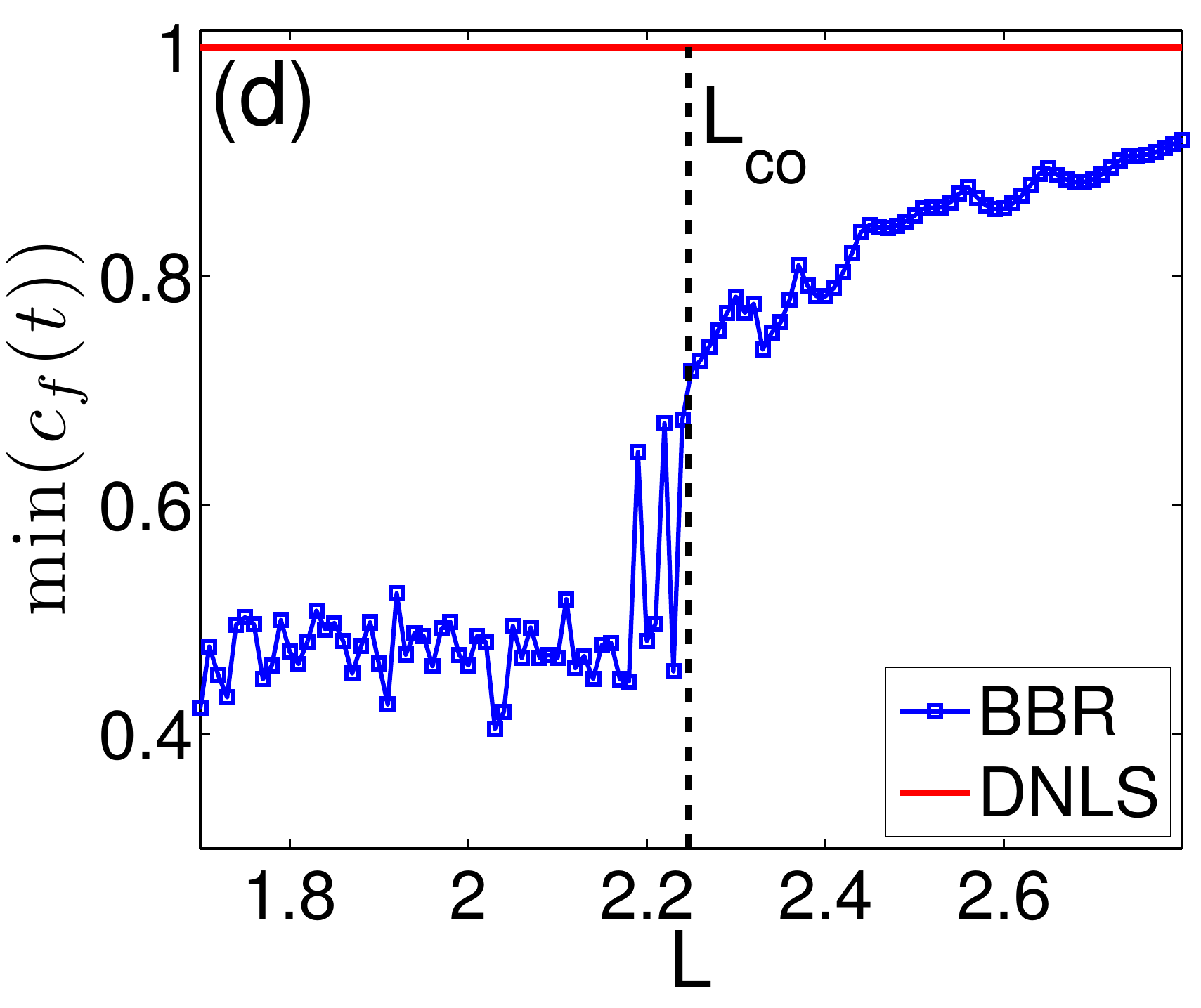}

\caption{
\textbf{(a)} The PN energy landscape $H_{\text{PN}}$ exhibits a bright DB (located in the top `eye' on the PN landscape at $N_2=0.902$ and $N_1=N_3=0.049$) and two degenerate saddle points that mark the boundary of stability of the DB. The color code shows the PN landscape (Eq.~(\ref{eq:Hpn})) for $L=L_{\text{co}}=2.2469$.
\textbf{(b)} The ST crossover is found at the crossing (dashed line)
of the total energy $E_0=L/2$ (red line) with the PN energy barrier $E_{\text{PN}}$
(Eq.~(\ref{eq:Epn}), green line). The corresponding energy at the crossing is the total energy of the saddle points shown in (a).
For $L\!\to\!\infty$, $E_0$ asymptotically approaches the total energy of the bright breather $E_b$ (blue line).
\textbf{(c-d)} Including quantum corrections
using the BBR method (blue line), the ST crossover exhibits a much
sharper transition compared to the mean-field result (red line), here
shown for $M=9$ sites.
We report the minimum number of atoms at the central site $\min(N_{5})$ and the condensate fraction $\min(c_f(t))$. 
The dashed line in (c-d) marks the crossover at $L_{\text{co}}$ for an infinite lattice.
Other parameters are
$\gamma=0.5$ Hz, $N=200$ atoms, $J=10$ Hz. The minima were determined
in the interval $t\in[0.25,0.5]\,$s. 
\label{fig:PN}}
\end{figure}

Let us first consider the dissipationless 
case (which belongs to type I) with the following initial
condition, where all atoms are located at site $c$, given by
\begin{equation}
\psi_{n}(t=0)=\delta_{nc}\,.\label{eq:initial_condition}
\end{equation}
In which range of the nonlinearity will the majority of the atomic population
stay self-trapped near site $c$ (resulting in the formation of
a DB)? In Fig.~\ref{fig:crossover} the evolution
of the particle density is shown. For $L=1.6$ (Fig.~\ref{fig:crossover}(a))
the particle density initially decays exponentially in time and then
populates the whole lattice evenly. In contrast, a completely different
behavior is observed for $L=2.4$ in Fig.~\ref{fig:crossover}(b),
where the initial condition relaxes into a ST state which is exponentially localized in space. A necessary condition
for ST is $L_{n}^{\text{local}}>L_{\text{co}}$,
where $L_{\text{co}}$ is the value of the nonlinearity at the {\em crossover}
that separates the diffusive from the ST regime. 
We define that ST is encountered, if $\min|\psi_{c}(t>T)|^{2}>a$ for large $T$,
which is independent of $T$ once a breather has formed. 
The value for $a$ can be estimated via the position of a saddle point (in the so-called Peierls-Nabarro energy landscape of a local trimer model, see below)
that dictates the stability of the DB \cite{Hennig:2010gy}, which is shown in Fig.~\ref{fig:PN}(a). In the limit $L\!\to\!\infty$ the saddle point is found analytically at $N_2=1/2$ \cite{Hennig:2010gy}, we therefore estimate $a=1/2$.
Starting with initial condition (\ref{eq:initial_condition}) and
choosing $T=1\,$s, the crossover from diffusion to ST
is numerically found to be at $L_{\text{co}}^{\text{num}}=2.2463$. 
Integration times were at least $10T$.

In the following, we will examine the observed ST crossover in detail,
for which we make use of a general concept called the PN energy barrier.
It is given by the energy difference $|E_{b}\lyxmathsym{\textminus}E_{e}|$,
where $E_{b}$ is the total energy of a DB centered at a single lattice site 
and $E_{e}$ is the energy of a more extended breather
centered between two lattice sites \cite{Kivshar:1993vf,Rumpf:2004en}.
The PN barrier is based on the notion, that due to continuity, the process of translating a localized object with energy $E_b$ from one lattice site to the adjacent one involves an intermediate state with different energy $E_e$.
%

We will connect the ST crossover to the stability of a DB. It has been shown that
the stability of a DB can be well-described via a reduced problem of only few degrees of freedom \cite{Flach:2008ud},
which reflects the fact that the breather is highly (exponentially)
localized. This `local Ansatz' has been further developed analytically
in a local trimer (which is a subsystem consisting of three sites) on the so-called PN energy landscape \cite{Hennig:2010gy}, which is defined by 
$H_{\text{PN}}=\max_{\delta\phi{}_{ij}}(H)$, 
with $\psi_n=\sqrt{N_n}\exp (i\phi_n)$ and $\delta\phi{}_{ij}=\phi_{i}-\phi_{j}$ 
\footnote{In ref.~\cite{Hennig:2010gy} $(-H_{\text{PN}})$ is called the \textit{lower} PN landscape due to the additional minus sign. A second energy landscape is obtained via $H_{\text{PN}}^\ast=\min_{\delta\phi{}_{ij}}(H)$, however, to study the ST crossover it is sufficient to consider only 
$H_{\text{PN}}$.}.
The PN landscape reads \cite{Hennig:2010gy}
\begin{equation}
H_{\text{PN}}=\frac{L}{2}(N_{1}^{2}+N_{2}^{2}+N_{3}^{2})+(\sqrt{N_{1}}+\sqrt{N_{3}})\sqrt{N_{2}}\,. \label{eq:Hpn}
\end{equation}

Figure \ref{fig:PN}(a) shows the PN landscape of the trimer
at the ST crossover. The bright DB, which is linearly stable \cite{Buonsante:2003dd}, is located in the top `eye' of the energy landscape.
The two saddle points just below $N_2=1/2$ (related to a migration of the DB from site $2$ to site $1$ and $3$ respectively) are connected to the PN barrier and the total energy threshold dictating the breather stability is given by \cite{Hennig:2010gy}
\begin{equation}
E_{\text{PN}}(L)=\frac{L}{4}+\frac{1}{2}+\frac{1}{4L}-\frac{1}{4L^{2}}+\frac{1}{4L^{3}}-\frac{9}{16L^{4}}+{\cal O}(\frac{1}{L^5})\,.\label{eq:Epn}
\end{equation}
The energy of a bright breather $E_{b}$ is a maximum of
the total energy $E$ of the trimer. As long as the total energy of
the local trimer $E_{\text{PN}}<E\le E_{b}$ is above the threshold,
a breather remains pinned to a lattice site. The total energy for the initial condition (\ref{eq:initial_condition}) reads $E(L)=L/2$, which can be seen directly 
from Eq.~(\ref{eq:hamiltonian}) as the energy is measured in units of the tunneling rate J (cf.~Eq.~(\ref{eq:dnls})).
Hence, the crossover $L_{\text{co}}$ is reached, when $E_{\text{PN}}$ (Eq.~(\ref{eq:Epn})) is equal to $L/2$, and we obtain
\begin{equation}
L_{\text{co}}^{5}-2L_{\text{co}}^{4}-L_{\text{co}}^{3}+L_{\text{co}}^{2}-L_{\text{co}}+\frac{9}{4}=0\,.\label{eq:polynom5}
\end{equation}
We find $L_{\text{co}}=2.2469$, in excellent agreement with the
numerical value. This result means that the ST crossover, which
is observed in a one-dimensional optical lattice, can be described
with high degree of accuracy by the PN barrier of a local trimer.
Given that the PN barrier describes the stability of self-trapped
states in a very broad context, we expect that the three different
types to obtain ST (static, dynamical and self-localized) in discrete
systems eventually are related to the PN barrier.

The general behavior near the ST crossover is depicted
in Fig.~\ref{fig:PN}(b). The PN barrier bends off the total energy
line for increasing $L>L_{\text{co}}$, which leads to a growing
area of stability (given by $E_{0}(L)>E_{\text{PN}}(L)$). In the
limit $L\!\to\!\infty$, the initial total energy $E_{0}$ (red line)
asymptotically approaches the total energy $E_{b}$ of the bright
breather (blue thick line)
\footnote{For $L\ll1$ the stability of a bright breather is not described by Eq.~(\ref{eq:Epn}) \cite{Hennig:2010gy}.}.
The exact breather energy $E_{b}(L)$, here for $M=101$ sites can be calculated numerically using standard methods (such as the anti-continuous limit \cite{Flach:2008ud,Marin:1996wv,Aubry:1997wo}), while we applied a different iterative approach  \cite{Proville:1999vb}.

\subsection{Generalized BBR method}
To study, how the ST crossover manifests beyond the mean-field description,
we use the Bogoliubov Backreaction (BBR) method \cite{Vardi:2001vw,Tikhonenkov:2007hm}, which 
includes higher-order correlation functions and allows a consistent calculation of the condensate fraction of the BEC. The BBR method has recently been generalized to the dissipative case \cite{Witthaut:2011kv,Trimborn:2011gy}, which is crucial for our study.
The generalized BBR method is especially useful if the many-body state is
close to, but not exactly equal to a pure BEC, in particular it 
accurately predicts the onset of a depletion of the condensate
mode \cite{Witthaut:2011kv}. 

We shortly review the main steps of the derivation of the generalized BBR method and point out its validity.
The coherent dynamics of ultracold atoms in deep optical lattices
is described by the Bose-Hubbard Hamiltonian
\bea
  \hat H = - J  \sum \nolimits_{j} \left( \hat a_{j+1}^{\dagger} \hat a_j +
                  \hat a_{j}^{\dagger} \hat a_{j+1} \right)
         + \frac{U}{2} \sum \nolimits_j 
           \hat a_{j}^{\dagger}  \hat a_{j}^{\dagger} 
               \hat a_{j}  \hat a_{j},
    \label{eqn-hami-bh}
\eea
where $\hat a_j$ and $\hat a_j^\dagger$ are the bosonic annihilation and 
creation operators, $J$ denotes the tunneling
rate between the wells and $U$ is the on-site interaction.
The BHH is obtained when the lattice is sufficiently deep, such that
the dynamics is restricted to the lowest Bloch band. We measure energy in frequency units by setting $\hbar = 1$.

To consider the quantum dynamics in presence of dissipation, we use a master
equation in Lindblad form \cite{Breuer:2002th}
\be 
  \dot{\hat \rho} =  -i [\hat H,\hat \rho]  + \LL \hat \rho.
  \label{eqn-master}
\ee
Localized particle loss and phase noise are described by the Liouvillians
\cite{Breuer:2002th}
\be
  \LL_{\rm loss} \hat \rho = -\frac{1}{2} \sum \nolimits_j \gamma_j \left(
     \hat a_{j}^{\dagger} \hat a_{j}  \hat \rho
     + \hat \rho \hat a_{j}^{\dagger} \hat a_{j} 
     - 2 \hat a_j \hat \rho \hat a_{j}^{\dagger}  \right),
\ee
\be     
    \LL_{\rm phase} \hat \rho = 
    -\frac{\kappa}{2} \sum \nolimits_j \hat n_j^2 \hat \rho + \hat \rho \hat n_j^2
       - 2 \hat n_j \hat \rho \hat n_j,
\ee
where $\gamma_j$ denotes the particle loss rate at site $j$ and
$\kappa$ is the strength of the phase noise.

In this article, we set $\kappa=0$, thus considering only particle loss.
For the purpose of generality, the terms resulting from phase noise
are included below.

We will first derive the mean-field equations from this -- so far exact -- approach. The higher order correlation functions that are the building block of the BBR method will then appear naturally.
We start from the single particle reduced density matrix (SPDM)
$\sigma_{jk} = \langle \hat a_j^\dagger \hat a_k \rangle
=\tr(\hat a_j^\dagger \hat a_k \hat \rho)$ 
\cite{Vardi:2001vw,Anglin:2001wq,Tikhonenkov:2007hm,Trimborn:2008uo}.
The equations of motion for $\sigma_{jk}$ are  obtained from the 
master equation (\ref{eqn-master})
\bea  
  i\frac{d}{dt} \sigma_{j,k} 
  &=&  \tr \left(\hat a_j^\dagger \hat a_k [\hat H, \hat \rho]
                  + i \hat a_j^\dagger \hat a_k   \LL \hat \rho \right) \nn \\
  &=&   -J \left( \sigma_{j,k+1} + \sigma_{j,k-1}
        - \sigma_{j+1,k} - \sigma_{j-1,k} \right) \nn \\
  && + U  \left( \sigma_{kk} \sigma_{jk} + \Delta_{kkjk} 
               - \sigma_{jj} \sigma_{jk} - \Delta_{jjjk} \right),  \nn \\ 
   &&   - i \frac{\gamma_j + \gamma_k}{2} \sigma_{j,k} 
     - i \kappa (1-\delta_{j,k}) \sigma_{j,k} ,
               \label{eqn-mf-final}
\eea
with the variances
$
   \Delta_{jk \ell m} = 
      \langle \hat a_{j}^\dagger \hat a_{k}  \hat a_{\ell}^\dagger \hat a_{m} \rangle
       - \langle \hat a_{j}^\dagger \hat a_{k} \rangle 
        \langle \hat a_{\ell}^\dagger \hat a_{m} \rangle .
$
In the mean-field limit $N \rightarrow \infty$ (where $UN$ remains finite),
one can neglect the variances $\Delta_{jk \ell m}$ in 
Eq.~(\ref{eqn-mf-final}) in order to obtain a closed set of 
evolution equations. This is the case for a pure BEC, as
the variances scale only linearly with the particle number $N$, 
while the products $\sigma_{jk} \sigma_{\ell m}$ scale as $N^2$.

To describe many-body effects such as quantum correlations and the depletion of
the condensate for large, but finite particle numbers, we explicitly take the variances $\Delta_{jk \ell m}$ into account. The time evolution of the variances $\Delta_{jk \ell m}$ includes six-point correlation functions $ \langle  \hat a_j^\dagger \hat a_m  \hat a_k^\dagger \hat a_n 
       \hat a_r^\dagger \hat a_s \rangle$. And the equations of motion for the six-point function then contain even higher correlation functions and so on.
In order to obtain a closed set of equations of motion, the higher-order (six-point) correlation functions are truncated as follows \cite{Tikhonenkov:2007hm}:
\bea
  &&   \langle  \hat a_j^\dagger \hat a_m  \hat a_k^\dagger \hat a_n 
       \hat a_r^\dagger \hat a_s \rangle \approx
    \langle  \hat a_j^\dagger \hat a_m  \hat a_k^\dagger \hat a_n \rangle
    \langle \hat a_r^\dagger \hat a_s \rangle \nn \\
  && \qquad \qquad  +  
     \langle  \hat a_j^\dagger \hat a_m  \hat a_r^\dagger \hat a_s \rangle
    \langle \hat a_k^\dagger \hat a_n \rangle
   +  \langle \hat a_k^\dagger \hat a_n \hat a_r^\dagger \hat a_s \rangle
       \langle \hat a_j^\dagger \hat a_m \rangle \nn \\
 && \qquad \qquad    - 2     \langle  \hat a_j^\dagger  \hat a_m  \rangle
        \langle  \hat a_k^\dagger \hat a_n \rangle
       \langle \hat a_r^\dagger \hat a_s \rangle.
\eea
Within this framework, we see that the mean-field approximation results from truncating the four-point correlation functions (and thus neglecting the variances $\Delta_{jk \ell m}$), while within the BBR approach the four-point functions are taken explicitly into account and the six-point functions are truncated.
With this ansatz we obtain the generalized BBR equations of motion (see \cite{Witthaut:2011kv,Trimborn:2011gy} for details).
The relative error induced by the truncation vanishes as $1/N^2$ with increasing particle number.
Close to a pure condensate, the BBR method thus provides a much more accurate description of the many-body dynamics than the simple mean-field approximation. 

\subsection{ST crossover beyond mean-field}
In Fig.~\ref{fig:PN}(c) the minimum remaining number of atoms (normalized
to $1$) at the central site are shown for $M=9$ sites and initial condition (\ref{eq:initial_condition}) using the generalized BBR method
and compared to the mean-field result. 
Boundary dissipation was applied in both cases, reducing reflections from the edges of the lattice.
The condensate fraction $c_f$ is the fraction of the number of condensed atoms and is given by the largest eigenvalue of the SPDM $\sigma_{j,k}$,
whereas the total number of atoms is given by the trace of
$\sigma_{j,k}$ \cite{Vardi:2001vw,Anglin:2001wq,Leggett:2001tp}.

The crossover at $L_{\text{co}}$ (dashed line in Fig.~\ref{fig:PN}(c-d)), which we have derived in Eq.~\ref{eq:polynom5}, is in excellent agreement with the BBR calculations. By including quantum corrections,
the ST crossover becomes much sharper which is also reflected by a
jump in the condensate fraction, see the blue curve in Fig.~\ref{fig:PN}(d), where we report $\min(c_f(t))$ for times $t\in[0.25,0.5]\,$s.
In contrast, the mean-field dynamics based upon the DNLS per se assumes
a pure BEC, i.e., $c_f=1$ (red line). 

While stable motion above
the crossover allows for long-time coherence, unstable motion below
the crossover leads to depletion of the condensate \cite{Castin:1997vx}.
A profound understanding of the ST crossover therefore might be
viable for controlled quantum state preparation using spatially
localized initial conditions, such as Eq.~(\ref{eq:initial_condition}).
%

\begin{figure}
\includegraphics[width=0.75\columnwidth]{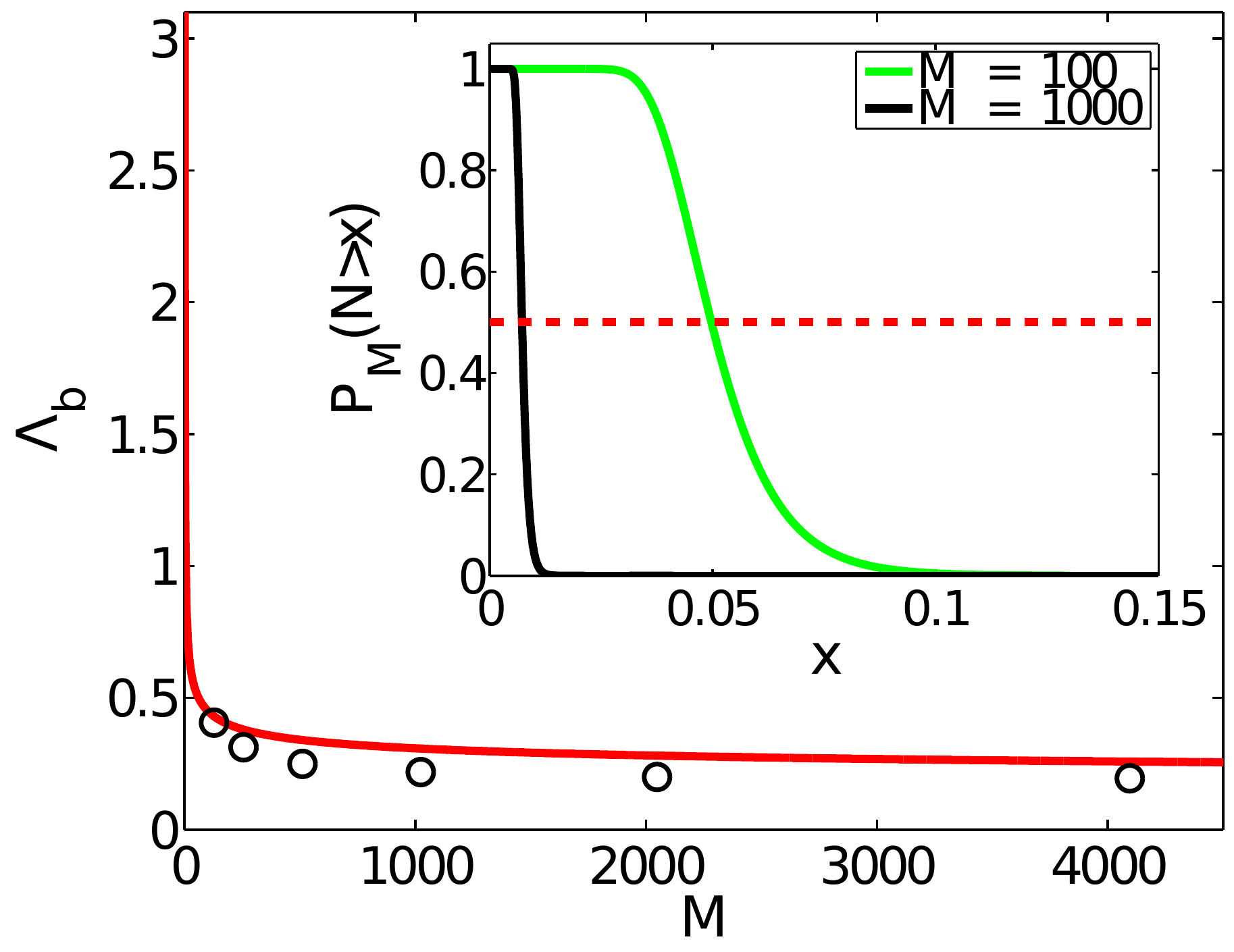}

\caption{The SL transition at $\Lambda_{b}$ (red line) given
by Eq.~(\ref{eq:Lambdac_final}) is an upper bound numerical
results. The data (black circles) was extracted for boundary dissipation rate $\gamma=0.2$ Hz from Fig.~$1$ in \cite{Ng:2009tu}, where a sharp drop of the `participation ratio'
clearly marks the SL transition. The inset shows the
probability for detecting a norm $x$ (Eq.~(\ref{eq:pofx}))
which exhibits a sharp transition and becomes a step function in the
limit $M\to\infty$.\label{fig:lc}}
\end{figure}

\section{Self-localization}
We now turn our focus to SL, where the dynamics finds
self-trapped states by itself in presence of weak boundary dissipation \cite{Livi:2006id,Franzosi:2007ds,Ng:2009tu,Franzosi:2011ft},
resembling a phase transition \cite{Ng:2009tu}. To consistently investigate
the dynamics in different lattices sizes $M$, we require the initial
density $\rho={\cal{N}}/M$ to be constant. Rescaling $L$
accordingly results in an effective nonlinearity
$\Lambda=L/M$ \cite{Ng:2009tu}. Starting with a homogeneous
initial condition with equal norm on all lattice sites and random
phases, the transition to SL has been observed at a
critical interaction strength $\Lambda_{b}$ for which we will derive
an explicit expression in the following. 
The condition for self-trapping reads
$L_{n}^{\text{local}} = L N_n  = \Lambda M N_{n}  > L_{\text{co}}$. 
The critical nonlinearity $\Lambda_{b}$ for the dynamical formation of a breather
is obtained for $L_{n}^{\text{local}}=L_{\text{co}}$, hence we find
\begin{equation}
\Lambda_{b}=\frac{L_{\text{co}}}  {MN_{m}}\,.\label{eq:Lambdac}
\end{equation}
As the only unknown quantity in Eq.~(\ref{eq:Lambdac}) is the maximum
single site norm $N_{m}$, calculating $\Lambda_{b}$ is reduced to
a very general question: 
What is the probability to find a site with norm larger than a given value N in the optical lattice?
In the diffusive regime, the probability distribution
of norms $x$ in the lattice is $w(x)=M\exp(-Mx)$ \cite{Ng:2009tu}.
The probability that the norm at a certain site is smaller than x
is
\begin{equation}
P(N<x)=\int_{0}^{x}w(x^{\prime})dx^{\prime}=1-e^{-Mx}\,.\label{eq:prob}
\end{equation}
Assuming that the populations at the $M$ sites are independent from
each other, the probability that at least one site has a norm larger
than $x$ reads
\begin{equation}
P_{M}(N>x)=1-[1-e^{-Mx}]^{M}\,,\label{eq:pofx}
\end{equation}
which approaches a step
function for $M\to\infty$ (see inset of Fig.~\ref{fig:lc}). Thus,
the largest norm that is found in the diffusive regime is given for large $M$ by $P_{M}(N>x)\approx1/2$
(red dashed line in Fig.~\ref{fig:lc}). Insertion into Eq.~(\ref{eq:pofx}) yields
\begin{equation}
N_{m}\equiv x=\ln[\frac{1}{1-(1/2)^{1/M}}]/M\,.\label{eq:Nn}
\end{equation}
With Eq.~\ref{eq:Lambdac} the SL transition is found to be at the critical nonlinearity
\begin{equation}
\Lambda_{b}=\frac{L_{\text{co}}} {\ln[\frac{1}{1-(1/2)^{1/M}}]}\,, \label{eq:Lambdac_final}
\end{equation}
which is shown in Fig.~\ref{fig:lc} (red line). 
As in deriving Eq.~(\ref{eq:Nn}) it was assumed
that the populations at the $M$ sites are independent,
we have effectively calculated an upper bound to $\Lambda_{b}$, in
excellent agreement with the numerical results in \cite{Ng:2009tu}
(shown as black circles in Fig.~\ref{fig:lc}).

\section{Experimental realization}

We expect that the ST crossover can be readily studied in present experiments \cite{Anker:2005vf,Gericke:2008tq,Wurtz:2009bh}. The initial condition (\ref{eq:initial_condition}) relates to a BEC cloud at
a single lattice site, while the interatomic interaction can be tuned via a Feshbach resonance.
In contrast, observing SL in optical lattices is more delicate.  A prerequisite to observe SL is dissipation.

While discrete breathers exist as well in Hamiltonian systems \cite{Flach:2008ud,Campbell:2004vz}, they become attractors of the dynamics in dissipative systems \cite{Flach:1998ul,Martnez:2003bw,MacKay:1998um} which is crucial for SL. Furthermore, local dissipation helps stabilizing once formed DBs by damping down phonons in the lattice.

Local dissipation has been realized with single site resolution using a focused electron beam \cite{Gericke:2008tq,Wurtz:2009bh}, while another possibility is to apply a microwave field to locally spin flip atoms inside the BEC \cite{Bloch:1999tj,Ottl:2005ho}.
The experiment, however, needs to allow for sufficient
propagation time so that SL can form, in the course of which chaotic
dynamics and dynamical instabilities typically lead to depletion of
the condensate \cite{Castin:1997vx,Fallani:2004hb,Bloch:2005uv}. A remedy could
be to reduce the timescale by considering lattices with few number of wells (as in Fig.~\ref{fig:sl2}) or to prepare an initial condition,
that has more than exponentially small probability for high norms. %

\section{Conclusion}
In conclusion, we analyzed the nature of SL
in optical lattices, which previously has been observed phenomenologically in several studies \cite{Livi:2006id,Franzosi:2007ds,Ng:2009tu,Franzosi:2011ft}, explaining recent numerical findings \cite{Ng:2009tu}. SL represents an alternative way to induce localization where the preparation of initial wave packets is not necessary.

Our results show that the SL transition at $\Lambda_{b}$
for which we derived an explicit estimate (Eq.~(\ref{eq:Lambdac_final}))
is based upon two constituent parts.
The first part is a ST crossover,
which we studied both within and beyond the mean-field description.
The second part is based on the probability, that the dynamics leads to a local energy above the PN energy barrier.

Given the simplicity of initial condition (\ref{eq:initial_condition}) used
to probe the ST crossover, we expect that the crossover
is not only experimentally readily accessible, but that its understanding
could also be vital in generating long-time coherent states, without
the need to fine-tune the initial state.

\begin{acknowledgments}
We thank David K.~Campbell, Dirk Witthaut, J\'{e}r\^{o}me Dorignac and Thomas Neff for useful discussions. We acknowledge financial support by the German Research Foundation (DFG, grant no.~HE 6312/1-1 and Forschergruppe 760). \vspace{0.1 mm}
\end{acknowledgments}
\bibliographystyle{apsrev}
\bibliography{}

\end{document}